\documentclass[journal]{IEEEtran}

\usepackage[english]{babel}
\usepackage{mathptmx}

\usepackage[letterpaper,top=2cm,bottom=2cm,left=3cm,right=3cm,marginparwidth=1.75cm]{geometry}

\usepackage{amsmath}
\usepackage{graphicx}
\usepackage[colorlinks=true, allcolors=blue]{hyperref}
\usepackage{subcaption}
\usepackage{float}
\usepackage[normalem]{ulem}
\usepackage{cite}
\usepackage{algorithm}
\usepackage{algpseudocode}
\usepackage[dvipsnames,table,xcdraw]{xcolor}

\begin{document}

\title{Hierarchical Localization for Integrated Sensing and Communication}

\author{\IEEEauthorblockN{Giorgos Stratidakis\IEEEauthorrefmark{1},
Sotiris Droulias, and \\
Angeliki Alexiou,~\IEEEmembership{Member,~IEEE}}\\
\IEEEauthorblockA{Department of Digital Systems, University of Piraeus, Piraeus 18534, Greece\\
\IEEEauthorrefmark{1}Corresponding author: Giorgos Stratidakis (email:giostrat@unipi.gr)}
}




\maketitle
\begin{abstract}
Localization is expected to play a significant role in future wireless networks as positioning and situational awareness, navigation and tracking, are integral parts of 6G usage scenarios. Nevertheless, in many cases localization requires extra equipment, which interferes with communications systems, while also requiring additional resources. On the other hand, high frequency and highly directional communications offer a new framework of improved resolution capabilities in the angular and range domains.
The implementation of integrated sensing and communications is being explored to unify the sensing and communications systems and promote a communicate-to-sense approach. To this end, a localization algorithm is presented that utilizes beam-forming and the emerging beam-focusing technique, to estimate the location of the receiver. The algorithm can be implemented with large antenna arrays, and large intelligent surfaces. The performance of the algorithm for static and mobile users is evaluated through Monte-Carlo simulations. The results are presented with the empirical CDF for both static and mobile users, and the probability of successful estimation for static users.

\end{abstract}
\section{Introduction}
Future wireless networks are expected to offer seamless connectivity and highly accurate sensing capabilities, and to achieve these goals higher frequency bands are being explored. In particular, the sub-terahertz (sub-THz) and the THz bands, offer higher data rates and sensing resolution than the lower frequencies \cite{Chaccour2022,Jiang2024,Balzer2022}. However, they are more limited in the transmission range due to the increased pathloss and are prone to outages due to blockage \cite{Galeote2024,Petrov2019,Stratidakis2020}. Furthermore, the probability of total blockage (i.e. the received power becoming zero) becomes higher with highly directional antennas.
In order to compensate for the reduced range, networks are becoming denser, and antennas are becoming more and more directional. The combination of highly directional antennas, blockage, and dense networks, makes sensing, and in particular localization, an necessary integral part of future wireless communications. More specifically, knowing and sharing the location of all nodes, static and mobile, allows for more efficient resource allocation, and increased power efficiency.

Typical localization schemes usually involve multiple nodes equipped with omni-directional antennas placed at different positions, and processes such as triangulation and trilateration \cite{Teoman2019,Yassin2017,Chen2022a}. In this case, to perform 3-dimensional localization, four nodes at different positions are required, while for 2-dimensional localization, three antennas are required \cite{Nagy2020}. With directional antennas, the required number of nodes is decreased by one for both 2-, and 3-dimensional localization \cite{Malhotra2005}. The reason for this is that directional antennas also estimate the general direction of the receiver (RX) in contrast to omni-directional antennas. Moreover, the location estimation becomes more accurate with antennas of higher directionality.
Localization, in many cases, is also performed with radars, and lidars, which operate independently of the communication system. 
As a result, in this case, the operation of simultaneous communications and localization requires additional hardware and resources.\\
\indent In the past few years, the concept of unifying sensing and communications was introduced \cite{Wei2023}. With integrated sensing and communication (ISAC), the same equipment and resources can be used to provide high data rate communications, and high resolution sensing, including localization, imaging and mapping \cite{Pin2021}, at the same time. For ISAC to work, sensing techniques that do not interfere with the communications aspect are required. Ideally, sensing should employ the existing communications systems, meaning a communicate-to-sense approach  which will simplify the implementation of the ISAC concept \cite{INSTINCT}. This means that in the case of communications with codebook-based beam-forming, the same codebooks will be used for sensing.
Today many localization algorithms have been proposed in works such as \cite{Huang2023,Lu2024,Que2023,El-Absi2018,Fan2021,Kanhere2020} that offer high accuracy. 
The authors in \cite{Huang2023}, explored the joint localization and environment mapping. Specifically, they employ multiple BSs and the multipath from multiple scatters, and RISs that act as one, to estimate the UE location, with two estimators. The first is a weighted least square estimator that exploits the non-line-of-sight (nLoS) components from the scatterers and RISs to perform localization and sensing jointly. The estimation is then refined with a second-stage estimators, that utilizes the output of the first one and nLoS path measurements. In \cite{Lu2024}, an hierarchical beam-training method for extremely large scale multiple-input-multiple-output (MIMO) systems in the near-field was presented. The authors design a polar-domain multi-resolution codebook for near-field beam-training. 
The authors in \cite{Que2023} propose an algorithm that performs localization and mapping simultaneously with one base station to estimate the location of the user equipment (UE) with an error lower than $40$ cm. In \cite{El-Absi2018}, a localization algorithm was presented that is  based on chipless RFID systems in the THz band. The algorithm utilizes dielectric resonator (DR)-lens tags to offer mm-level accuracy in indoors scenarios. The authors in \cite{Fan2021} proposed a localization scheme where three transmitters (TXs) cooperate for the UE, with the help of a new deep learning approach, to obtain its location with average localization error $\sim24$ cm. Furthermore, the broad spectrum ($5$ GHz), and the THz band they use, along with the distance-adaptive multi-path effect this band offers, results in great temporal and spatial resolution, which is ideal for localization purposes. In \cite{Kanhere2020}, the authors utilize the multipath components of transmission in $28$ GHz with a horn antenna of $30^o$ 3dB beam-width at both the TX and the RX, and $140$ GHz with a horn antenna of $8^o$ 3dB beam-width at the TX and the RX. They achieve a mean localization error of $5.72$ cm at $28$ GHz, and $6.29$ cm at $140$ GHz. 

\indent In the above-mentioned works, the proposed approaches in some cases require additional equipment to achieve high accuracy, such as multiple TXs or DR-lens tags, which limit their applicability to use cases where this equipment can be placed. Furthermore, approaches that take advantage of multipath for the location estimation may find weak applicability in high frequencies where the multipath is poor, and is further suppressed with highly directional antennas (e.g. pencil beams), which are used to counteract the increased pathloss. Moreover, most of the aforementioned works are not easily implemented in ISAC systems.

In this work, an hierarchical localization algorithm is presented that is suitable for use with large antenna arrays (LAAs) and large intelligent surfaces (LISs) and offers accuracy within the limits for the use cases shown in \cite{Wymeersch2021}. The algorithm is tailored to take advantage of existing hardware and techniques, in order to estimate the location of the RX, which makes it suitable for ISAC. The algorithm consists of two phases. In Phase 1, the TX finds the direction of the RX with hierarchical beam-training and after it reaches sufficient accuracy switches to Phase 2, which is hierarchical ranging, to estimate the distance with beam-focusing. After estimating both the direction and distance of the RX, the calculation of the location is straightforward. The main advantage of this approach is its simplicity; hierarchical ranging is the equivalent of hierarchical beam-training, which is known for its low complexity. 
The main contributions of this paper are as follows
\begin{itemize}
    \item An analytical way to generate focused beams with the desired focal area size is presented.
    \item A localization framework for the communicate-to-sense concept is proposed.
    \item A localization algorithm is proposed that is composed of an hierarchical beam-training and a novel hierarchical ranging algorithm.
    \item The proposed hierarchical ranging algorithm is the equivalent of the hierarchical beam-training approach in range. The method for the codebook generation of the beam-focusing codebook is also presented. 
    \item A simple approach for the beam-forming codebook is proposed to make the algorithm more robust to additive white Gaussian noise (AWGN).
\end{itemize}

\section{Beam-focusing-based localization}\label{s:2}
\begin{figure}
\centering
\includegraphics[width=\linewidth]{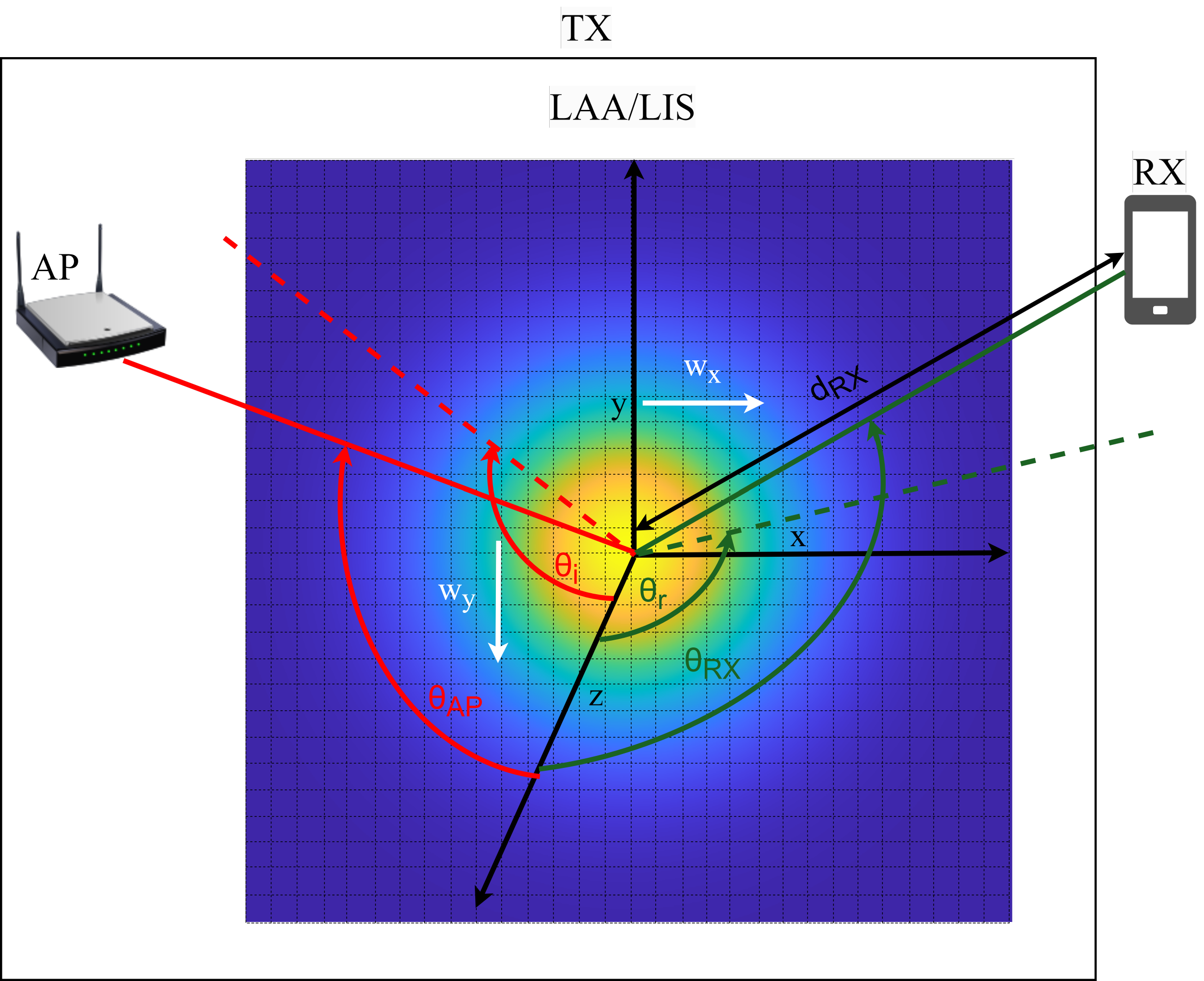}
\caption{System model. The TX (LAA or AP with LIS) performs beam-forming and beam-focusing towards the RX. The beam footprint, depicted with the density plot, is entirely contained in the TX panel.}
\label{fig:fig1}
\end{figure}

Beam-focusing-based localization takes advantage of the unique property of focused beams to concentrate the power within small areas, commonly referred to as \textit{focal areas}. The ability to concentrate the power of a beam to the desired location is considered fundamental in works that consider LAAs and LISs as the area of interest tends to be in the near-field. Moreover, in some works beam-focusing replaces beam-forming in the near-field \cite{Rao2024,Wu2024}. 
The TX generates multiple focused beams towards multiple target locations and, hence, the focal area of the beam that is closer to the RX provides the maximum power to the RX, and signifies its position. The TX can be an LAA that generates the focused beam towards the RX. Alternatively, the TX can be an LIS, which is illuminated by an access point (AP), acting as a secondary radiating element that reshapes the incident beam into a focused beam towards the RX. The system model is shown in Fig. \ref{fig:fig1}, where the footprint of the radiated beam at the TX is depicted with a density plot. In order for beam-focusing-based localization to work, the size of the focal areas must be known, and ideally be controllable. In this section, the equations to adjust the size of the focal areas are derived.
In \cite{Droulias2024} it was shown that for beam-focusing taking place on the $xz$-plane with a Gaussian footprint contained entirely in the TX panel, as is our case here, the power density distribution of the focused beam is given by
\begin{align}
\begin{split}
    S_r & (x,z) = \frac{2P_t}{\pi w_x w_y}\frac{1}{\sqrt{\left(1-\frac{z}{f_0 \cos{\theta_r}}\right)^2+(\frac{2z}{k w_y^2 \cos{\theta_r}})^2}}\times\\
    & \frac{1}{\sqrt{\left(1-\frac{z}{f_0 \cos{\theta_r}}\right)^2+\left(\frac{2z}{k w_x^2 \cos{\theta_r}}\right)^2 \left(\frac{1}{\cos{\theta_r}}\right)^4}}\times\\
    & \exp{\left[-\frac{2(x\cos{\theta_r}- z\sin{\theta_r})^2}{w_{x}^2 \left( \left(\cos{\theta_r}- \frac{z}{f_0} \right)^2+ \left(\frac{2z}{k w_x^2 \cos{\theta_r^2}}\right)^2\right)}\right]},
\end{split}
\label{Eq:Eq_Pr}
\end{align}
where $P_t$ is the transmitted power, $w_x$, and $w_y$ are the horizontal and vertical footprint radii, $\theta_r$ is the angle of the beam's direction on the $xz$-plane, $f_0$ is the intended focal distance, and $k$ is the wavenumber. In the case of LISs, $w_x=\frac{w_y}{\cos(\theta_i)}$, where $\theta_i$ is the angle of the incident beam (see Fig. \ref{fig:fig1}). Note that, when the focal distance is in the far-field of the beam, the TX essentially performs beam-forming. This becomes apparent by setting $f_0 \rightarrow \infty$.
The received power at the RX using Eq. (\ref{Eq:Eq_Pr}) is given by
\begin{align}
 P_r= S_r A_r,
 \label{Eq:Pr}
\end{align}
where $A_r= G_r \lambda^2 /4\pi$, and $G_r$ are the effective aperture and the antenna gain of the receiver, and $\lambda$ is the wavelength. 

As was shown in \cite{Droulias2022,Droulias2024}, for a certain $f_0$ along $\theta_r=0$, and horizontal and vertical footprint radii $w_x=w_y\equiv w$ the maximum power occurs at focal distance
\begin{align}
    d_f= \frac{f_0}{(f_0/z_R)^2+ 1},
    \label{Eq:Eq_d_f}
\end{align}
where $z_R=k w^2/2$ is the Rayleigh length, and $k$ is the wavenumber. For $z_R \gg f_0$, i.e. for relatively large footprints, $d_f\approx f_0$. The focal distance $d_f$ marks the center of an ellipse, with its major axis along the beam direction, on which the power is constant \cite{Wu2020}. Using Eq. (\ref{Eq:Pr}) the major radius of the 3dB focal area along $\theta_r=0$ is calculated as 
\begin{align}
    r_{max}= \frac{f_0}{(f_0/z_R)^2+ 1} \frac{f_0}{z_R}.
    \label{Eq:Eq_r_max}
\end{align}
The position and size of the 3dB elliptical focal areas are the key elements for partitioning the radial distance to achieve the desired localization resolution, which is determined by the size of the ellipse along the TX-RX direction, i.e. the major radius $r_{max}$. Note that $d_f$ and $r_{max}$ are controlled entirely by $f_0$ and $w$ (the latter through $z_R$) and, hence, the radial distance can be divided into as many ellipses as desired. According to Eq. (\ref{Eq:Eq_r_max}), for a fixed $w$ the size of the focal area increases with increasing $f_0$, and, hence, to form ellipses of equal size that guarantee uniform resolution, it is required that each beam is formed with different footprint. For $f_0\ll z_R$, the footprint practically changes proportionally to $f_0$ for
fixed $r_{max}$ (see Eq. (\ref{Eq:Eq_r_max})). 
By solving Eqs. (\ref{Eq:Eq_d_f}) and (\ref{Eq:Eq_r_max}) for $f_0$, the intended focal distance for a focal area of focal distance, $d_f$, and major radius, $r_{max}$, is
\begin{align}
    f_0= d_f \left(\left(\frac{r_{max}}{d_f}\right)^2+1\right).
    \label{Eq:f_distance}
\end{align}
Moreover, by solving for $w_x$, the radius of the beam footprint on the TX for the same focal area is
\begin{align}
    w_x= \sqrt{2 \frac{r_{max}^2+ d_f^2}{k r_{max}}}.
    \label{Eq:f_radius}
\end{align}
Eqs. (\ref{Eq:f_distance}) and (\ref{Eq:f_radius}) are used with Eq. (\ref{Eq:Eq_Pr}) to generate focal areas of desired major radius and focal distance, in order to control the resolution of a focusing-based localization algorithm.


\section{Hierarchical codebook-based exhaustive search}
\begin{figure}
\centering
\includegraphics[width=\linewidth]{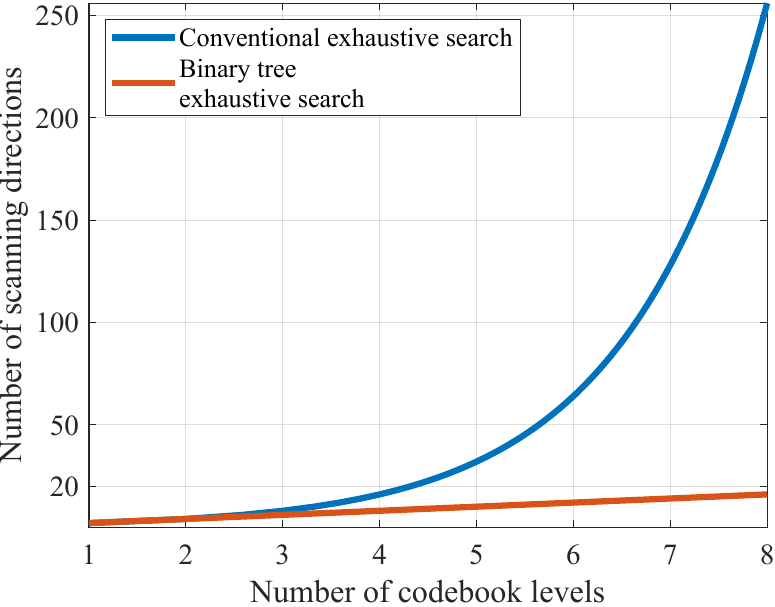}
\caption{Comparison between the conventional exhaustive search and the binary-tree-structured codebook-based exhaustive search used in this work in terms of directions to scan for different numbers of antenna elements. Conventional exhaustive search uses all codewords of only the last codebook level, while the binary tree exhaustive search uses $2$ codewords in each level.}
\label{fig:fig22}
\end{figure}

In this work, two binary-tree-structured codebooks are employed for localization, one for the direction estimation, and one for ranging. A codebook is a matrix that contains the weights for each antenna element of the LAA or LIS that are required to form the desired beams. In beam-training, codebooks are used to form beams at orthogonal directions that cover the entire area of interest. 
Since the size of a codebook is finite, the number of directions towards which the beams can be formed, is also finite. In this case misalignment is the result of the difference between the direction of the predefined beam and the direction of the RX.
The use of codebooks is chosen to reduce the number of pilots used for searching by reducing the number of directions to scan, as exhaustive search in a continuous fashion becomes highly time and resource consuming, especially with large arrays (LAAs or LISs) that form extremely narrow beams. Hierarchical codebooks include beams of different beam-widths. Each beam of one level corresponds to a number of beams of the next level, further simplifying the exhaustive search. Binary tree-structured codebooks are a special case of the hierarchical codebook, where, each beam corresponds to two beams in the next level. The use of binary tree-structured codebook enables exhaustive search with substantially less pilots in comparison to conventional exhaustive search \cite{Stratidakis2024}, as shown in Fig. \ref{fig:fig22}. Specifically, the conventional exhaustive search uses $2^{K}$ pilots, where $K$ is the number of codebook levels, to find the RX. With a binary tree-structure codebook, the exhaustive search algorithm uses $2$ pilots from each level for a total of $2K$ pilots. As a result the difference between the two methods increases with increasing number of antenna elements. After estimating which codeword, $cd$, in one level corresponds to the beam or focal area that is closest to the direction or distance of the RX, the codewords of the next level that correspond to this codeword can be calculated as $[2cd-1, 2cd]$.

\section{LAA/LIS-assisted Hierarchical localization algorithm}\label{s:sec4}

\begin{figure}
\centering
\includegraphics[width=\linewidth]{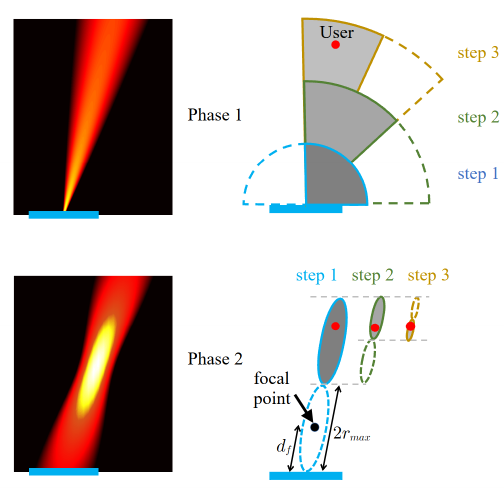}
\caption{Concept of the proposed localization algorithm. In Phase 1 the algorithm estimates the direction of the RX with beam-forming, and in Phase 2 the distance with beam-focusing. In the bottom left panel, the yellow coloured area marks the 3dB focal area.}
\label{fig:fig2}
\end{figure}

\begin{algorithm}
\caption{Localization algorithm.}{\label{alg:loc}}
\textbf{Beam-training with beam-forming}
\begin{algorithmic}
    \State $l_t=1$: Current beam-forming codebook
    \State $L_t$: Last beam-forming codebook level
    \State $S_t=[1, 2^{l_t}]$: codewords to use
    \While($l_t<L_t$)
    \State Scan the area using $S_t$ codewords of the codebook level $l_t$.
    \State Estimate the received power at the RX with each beam according to Eq. (\ref{Eq:Eq_Pr})
    \State Find the beam with the highest power at the RX.
    \State Estimate the direction of the RX, $\theta_{RX}$.
    \State Find codeword, $c_t$, of the estimation 
    \State Use codewords $S_t= [2c_t-1, 2c_t]$ of the higher codebook level to scan
    \State $l_t=l_t+1$
    \EndWhile
    \end{algorithmic}

\textbf{Ranging with beam-focusing}
\begin{algorithmic}
    \State $l_r=1$: Current beam-focusing codebook level
    \State $L_r$: Last beam-focusing codebook level
    \State $S_r=[1, 2^{l_r}]$: codewords to use
    \While
    \State Scan the focal areas generated by the $S_r$ codewords
    \State Estimate the received power at the RX with each focal area according to Eq. (\ref{Eq:Eq_Pr})
    \State Estimate the focal area with the highest received power the RX.
    \State Estimate the TX-RX distance, $d_{RX}$.
    \State Find the codeword, $c_r$ of the estimation
    \State Use codewords $S_r= [2c_r-1, 2c_r]$ of the higher codebook level to scan
    \State $l_r=l_r+1$
    \EndWhile
    \State Estimate the RX location as $[d_{RX} \sin(\theta_{RX}), d_{RX} \cos(\theta_{RX})]$ 
\end{algorithmic}
\end{algorithm}

Exhaustive search with focused beams can be very time consuming depending on the area of interest and the size of the focal areas. Therefore, a more efficient approach is required. The proposed localization algorithm in this work consists of two phases, beam-training and ranging, as shown in Fig. \ref{fig:fig2}, and outlined in \textbf{Algorithm \ref{alg:loc}}. Phase 1 utilizes a binary tree-structured codebook to divide the area of interest into smaller direction-based areas with beam-forming, and Phase 2 utilizes a binary tree structured codebook to divide the area of interest into distance-based parts with beam-focusing along the direction specified by Phase 1. 
The purpose of Phases 1 and 2 is to estimate the direction and distance, respectively, of the RX with controllable resolution.

\subsection{Phase 1: Beam-training}
The beam-training process in this work follows the binary tree search, which requires a beam-forming codebook, of $L_t=log_2 N$ levels, where $N$ is the total number of elements of an antenna array along the x-direction, where beam-steering takes place.
In this work, the beams provided by the codebook in \cite{Sayeed2015} were used to fill the levels of a new hierarchical codebook, which is here referred to as beam-forming (Bfr) codebook. 
The number of active elements is adjusted through the footprint radius. 
For example, the equivalent footprint radii for a $N\times N$ antenna array used in \cite{Sayeed2015} is here $w_x\approx\frac{N}{2}d_x$, $w_y\approx\frac{N}{2}d_y$ where $d_x$, and $d_y$ are the horizontal and vertical dimensions of each element.
The beam-training process is as follows. The TX generates beams with the two codewords of the first level of the Bfr codebook. The RX estimates the received power with both beams and sends the information to the TX. The TX estimates the direction of the RX based on the beam that offers the highest power at the RX. Then the TX generates beams based on the two codewords of the next level, that correspond to the codeword of the previous level and estimate the direction of the RX. This process is repeated until the highest codebook level is reached. The number of codewords in each level, is $2^{l_t}$, where $l_t=1,2,...,L_t$ is the current codebook level. With this process, the exhaustive search is carried out with $2L_t$ codewords, instead of $2^{L_t}$ which would be the case if only the highest codebook level was used in an exhaustive search.

\subsection{Phase 2: Ranging}
The ranging process also follows the binary tree structure. The first step is to generate a beam-focusing (Bfc) codebook, of $L_r$ levels that is used to form the desired focal areas. In order to do that, assuming that the maximum distance in the area of interest is $d_0$, the area of interest is divided in two smaller areas. Then each of the new areas is again divided in two smaller areas. This is repeated until the desired resolution is reached. The resolution in this phase is determined by $r_{max}$, the major radius of the focal area. Similar to the beam-forming codebook, the number of focal areas per codebook level, is equal to $2^{l_r}$, where $l_r=1,2,...,L_r$ is the current codebook level. For example, in the first codebook level the weights for two focal areas are stored, in the second, the weights for four focal areas, etc. 
\begin{figure*}[t]
\centering
\includegraphics[width=\linewidth]{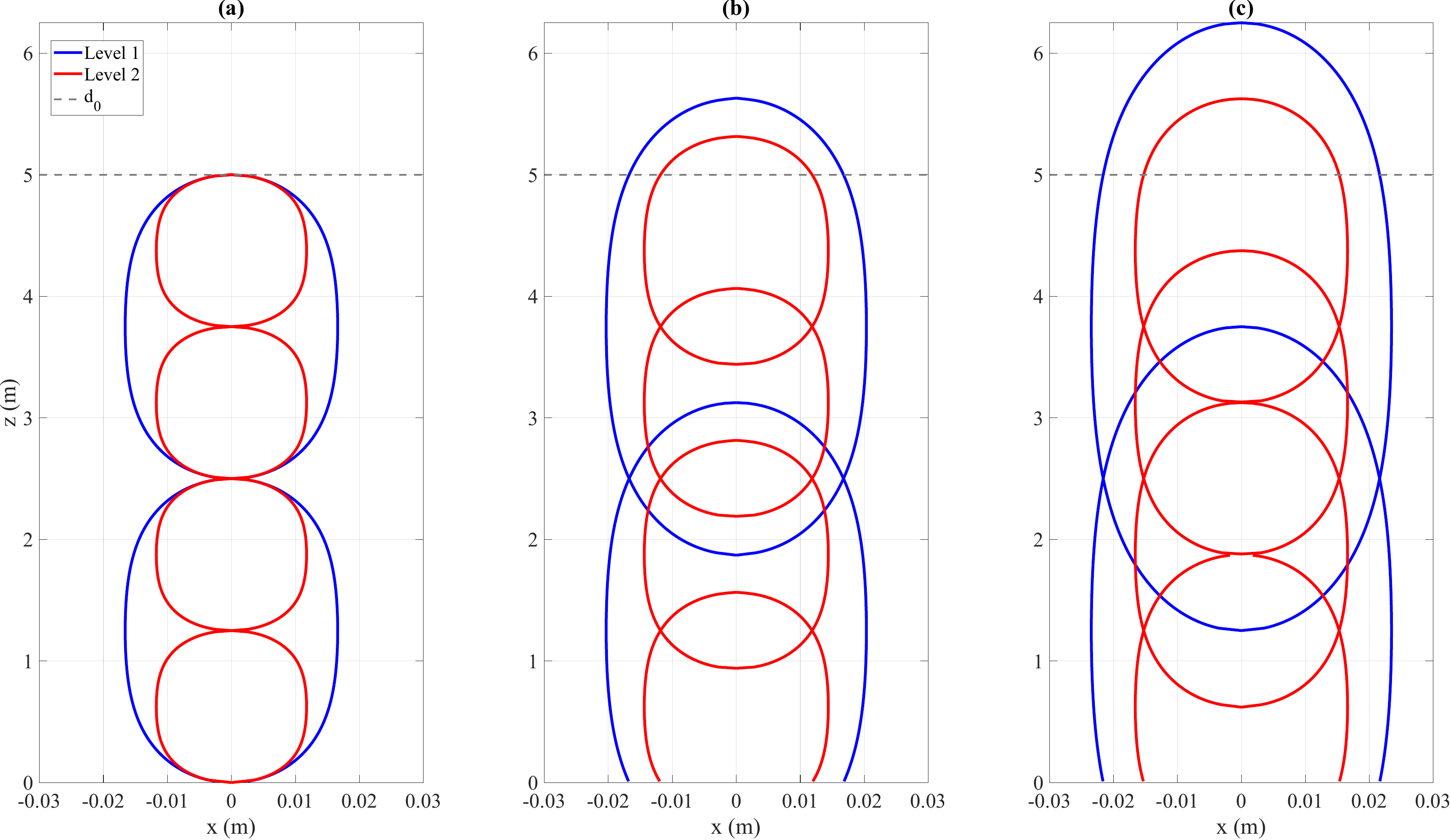}
\caption{Focal areas of a codebook with two levels for $d_0=5$ m, with (a) $\alpha=0.25$, (b) $\alpha=0.375$, and (c) $\alpha=0.5$.}
\label{fig:fig3}
\end{figure*}
The focal distances of the codebook can be calculated as 
\begin{align}
    d_f(l_r,i)= \frac{d_f(1,1)}{2^{l_r-1}} m_i,
    \label{Eq:Eq.7}
\end{align}
where $d_f(1,1)= d_0/4$ is the focal distance for the first codeword in the first codebook level, $d_0$ is the maximum distance from the TX in the area of interest, $i= 1,2,...,2^{l_r}$ is the codeword index. Furthermore, $m_i=1,3,5,...,M$ is a set of $2^{l_r}$ numbers that includes all the odd numbers from $1$ to $M= 2^{l_r+1}-1$.
The major radius of the focal areas can be calculated as 
\begin{align}
    r_{max}(l_r)= \alpha\ \frac{d_0}{2^{l_r-1}}, 
    \label{Eq:Eq.8}
\end{align}
where $\alpha\in[0.25, 0.5]$ is the focal area overlapping coefficient. For $\alpha\leq 0.25$, adjacent focal areas are formed with zero ovelap. At $\alpha=0.25$ the two ellipses meet at a single point, as shown in Fig. \ref{fig:fig3}(a). Lower values for $\alpha$ leave gaps between the focal areas and higher values increase the overlap between the focal areas, as demonstrated Fig. \ref{fig:fig3}(b), where $\alpha = 0.375$. The value $\alpha=0.5$ marks the limit, after which more than two adjacent focal areas overlap, as shown in Fig. \ref{fig:fig3}(c). If the RX belongs to two or more focal areas at the same time, the algorithm can choose either one and continue refining the estimated TX-RX distance. However, with larger values of $\alpha$, that make more than two focal areas to overlap, the focal areas become unnecessarily large as the resolution of each focal area is reduced substantially. For focal areas with $\theta_r\neq 0$, $\theta_r$ must follow the directions of the beam-forming codebook in Phase 1, with the same $d_f$, and $r_{max}$. To generate the focal areas for the codebook, the output of Eqs. \eqref{Eq:Eq.7}, and \eqref{Eq:Eq.8} is inserted to Eqs. \eqref{Eq:f_distance}, and \eqref{Eq:f_radius}.

\section{Performance assessment}
\begin{figure}
\centering
\includegraphics[width=\linewidth]{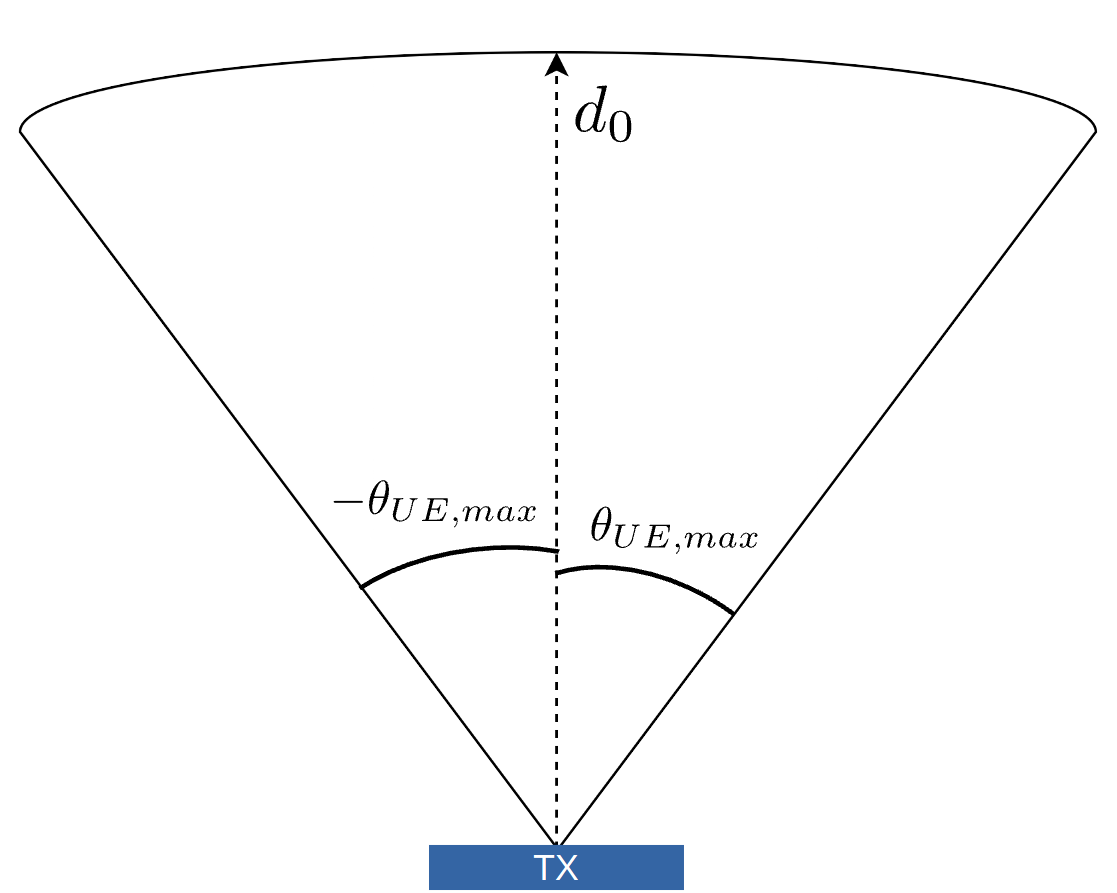}
\caption{Area of interest for scenarios 1, and 2 in the simulations. In both scenarios, $\theta_{UE,max}=25^o$. In scenario 1, $d_0=5$ m, and in scenario 2, $d_0=10$ m.}
\label{fig:figsc}
\end{figure}

\begin{table}[b]
\begin{tabular}{|l|l|l|l|}
\hline
codebook level & $d_0= 5$ m & $d_0= 10$ m\\ \hline
1        & 1.5 m     & 3 m        \\ \hline
2        & 0.75 m    & 1.5 m      \\ \hline
3        & 0.375 m   & 0.75 m     \\ \hline
4        & 0.187 m   & 0.375 m    \\ \hline
5        & 0.093 m   & 0.187 m    \\ \hline
6        & 0.046 m   & 0.093 m    \\ \hline
7        & 0.023 m   & 0.046 m    \\ \hline
8        & 0.011 m   & 0.023 m    \\ \hline
\end{tabular}
\caption{Resolution, $r_{max}$, of each beam-focusing codebook level for $d_0=5$ m, and $10$ m.}
\label{tab:tab1}
\end{table}

The performance of the proposed algorithm is evaluated through Monte-Carlo simulations of $10^6$ iterations (RX locations) in two scenarios. In both scenarios, the RX are placed randomly in an area that extends $\theta_{UE,max}= 25^o$ to the left and right of the TX broadside, starting at $10$ cm from the surface that the TX is placed at. The locations follow the uniform distribution in both angle and distance, and the shape of the area of interest is shown in Fig. \ref{fig:figsc}. 
In scenario 1, the maximum distance is $d_0= 5$ m from the center of the TX, and in scenario 2 it is $d_0= 10$ m. 
The central frequency is $150$ GHz, the spacing between the array elements are $d_x=d_y=\lambda/2$, the transmitted power is $30$ dBm, the antenna gain of the receiver is $G_r=20$ dB, the LIS size is much larger than the footprint, and the ellipses are formed with $\alpha=0.3$. 
In Table \ref{tab:tab1}, the resolution, $r_{max}$ of each beam-focusing codebook level is presented for the maximum distances of the two scenarios. 
In Phase 1, the beams of the beamforming codebook in level 1 span the entire ($-90^o$, $+90^o$) range. Here, because the area of interest is narrower, Phase 1 starts from the $4^{th}$ codebook level, where the beams are narrower and only the beams that cover the angular range of interest are used.
As the proposed localization is hierarchical, in Phase 1, the TX utilizes all codebook levels from the $4^{th}$ level and up of the beam-forming codebook using the binary tree approach to find the direction of the RX. In Phase 2, the TX uses all codebook levels of the beam-focusing codebook in the direction estimated in Phase 1 to find the TX-RX distance. 
In both scenarios, the beam-forming codebook in Phase 1 is composed of $10$ levels, and the beam-focusing codebook in Phase 2 of $6$ levels. 

\begin{figure}
\centering
\includegraphics[width=\linewidth]{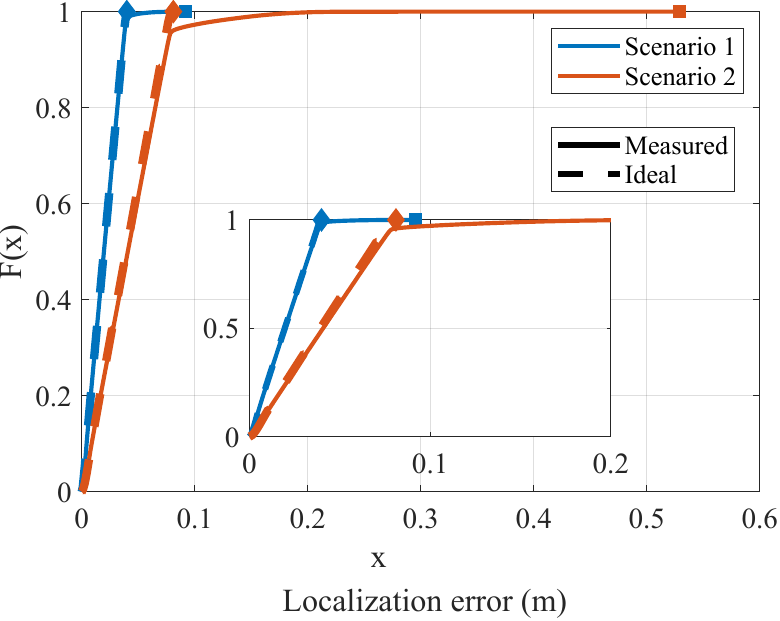}
\caption{Comparison between the ideal and measured performances of the algorithm in both scenarios. The squares mark the maximum measured error of the algorithm, and the rhombi the maximum error when the algorithm performs ideally.}
\label{fig:fig4}
\end{figure}

In Fig. \ref{fig:fig4}, the empirical CDF of the localization error is presented for the two scenarios, along with the CDF of the ideal performance of the algorithm in each scenario. The localization error is defined as the distance between the actual and the estimated location of the RX.
In the ideal case, the correct beam is always chosen from the Bfr codebook that corresponds to the direction of the RX in Phase 1, and the correct focal area from the Bfc codebook that corresponds to the TX-RX distance in Phase 2. The localization error derives from the fact that the RX position can be anywhere within the narrowest beam in Phase 1, and the smallest ellipse, in Phase 2. 
The same holds in the measured case. However, there are two reasons that can increase the localization errors with respect to the ideal case. 
The first reason is that in Phase 1 the RX direction uncertainty is in the order of the beamwidth, and is more pronounced for low codebook levels. As a result, the direction along which the focal areas are formed in Phase 2 may have an offset, leading to localization errors, especially for large RX distances.
The second reason is that there are instances where the RX position is estimated to be in the wrong ellipse, consequently increasing the localization error. The reason is that the definition of the focal areas are based on the analytical expressions (\ref{Eq:f_distance}),(\ref{Eq:f_radius}), which are exact for $\theta_r=0$, i.e. for RXs residing exactly in front of the TX. As was shown in \cite{Droulias2024}, the focal areas slightly change with $\theta_r\neq 0$. While this effect is weak for the small focal areas of the higher codebook levels, it is more pronounced in the lower levels, and any error that occurs there is carried over to the higher levels. Although this increases the average localization error, it is still low and comparable with other works that employ a single TX for localization.
The empirical CDF stops at the maximum error value that is estimated from the simulations. Therefore, this CDF also shows the overall maximum error for all RX locations, along with the maximum error for 99.9\% of the RX locations. In scenario 1, the overall maximum error is $9.2$ cm, the 99.9\% maximum error is $6.6$ cm, and the average localization error is $1.97$ cm, while in the scenario 2, they are $53$ cm, $20$ cm, and $4.2$ cm respectively. In the ideal case of scenario 1, the overall maximum error is $4$ cm, the 99.9\% maximum error is $3.9$ cm, and the average localization error is $1.95$ cm, while in scenario 2 they are $8.1$ cm, $7.9$ cm, and $4$ cm, respectively.  

%
%
It is observed that in scenario 2, the localization error is double that in scenario 1. The reason is that in scenario 2, the maximum distance $d_0$ is double that in scenario 1, but the same codebook level is reached, which offers different resolution for the two scenarios. Specifically, the $6^{th}$ level in scenario 2 offers half the resolution of the same level in scenario 1. Therefore, if an additional beam-focusing codebook level is used in Phase 2 of scenario 2, the same resolution as in scenario 1 can be achieved as shown in Table \ref{tab:tab1}.
%

\begin{figure}
\centering
\includegraphics[width=\linewidth]{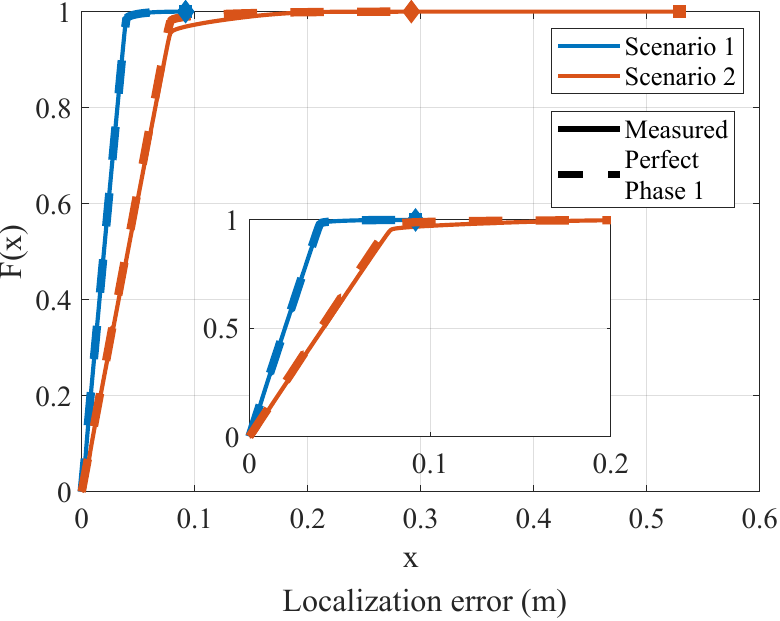}
\caption{Performance of the algorithm with perfect Phase 1.}
\label{fig:fig55}
\end{figure}

As previously stated, a part of the localization errors is the result of the direction uncertainty in Phase 1. The effect of Phase 1 in the localization errors is shown in Fig. \ref{fig:fig55}, where the empirical CDF of the localization error is presented for the measured case and the case with perfect Phase 1. Perfect Phase 1 means that in Phase 1 of the algorithm, the exact direction of the RX is known. By comparing the results of the two cases, the magnitude of the effect of Phase 1 in the localization errors can be extrapolated. In Scenario 1, where the maximum distance is $5$ m, the CDFs of the two cases are identical. However, in Scenario 2, where the maximum distance is $10$ m, the two CDFs are different. The most significant change is the lower maximum localization error when assuming perfect Phase 1, which in this case is $\sim22$ cm. There are two reasons for this. The first reason is that when the direction of the focal areas and the RX are not identical, a focal area different to the one closest to the RX will offer more power to the RX, and will be selected to serve the UE. This happens because the power outside the 3dB focal area, and along its minor radius decreases drastically, while the power along the major radius decreases at a slower rate, as shown in Fig. \ref{fig:fig2}.
The second reason is that with longer maximum distances, the focal areas of the same level become larger, which in combination to the first reason can further increase the localization errors.

\begin{figure}
\centering
\includegraphics[width=\linewidth]{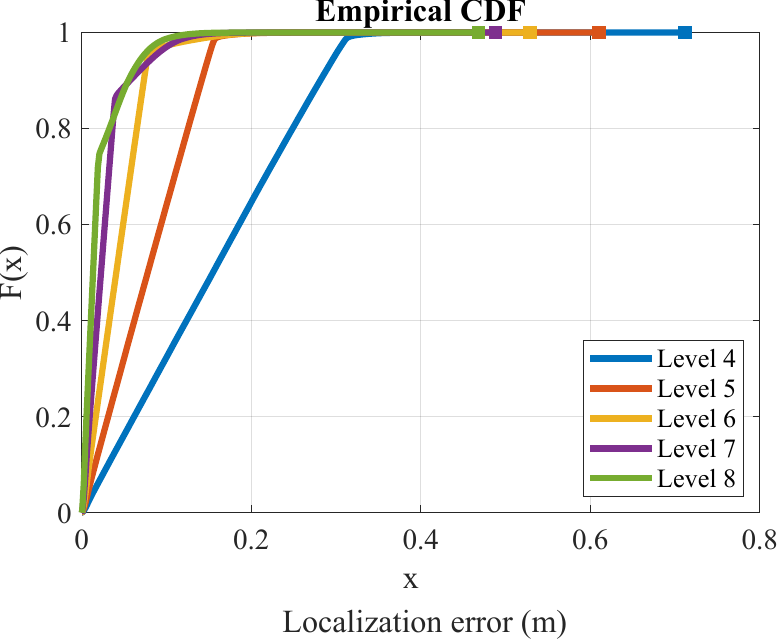}
\caption{Performance of the algorithm in Phase 2 at different beam-focusing codebook levels.}
\label{fig:fig8}
\end{figure}

\begin{figure*}
\centering
\includegraphics[width=\linewidth]{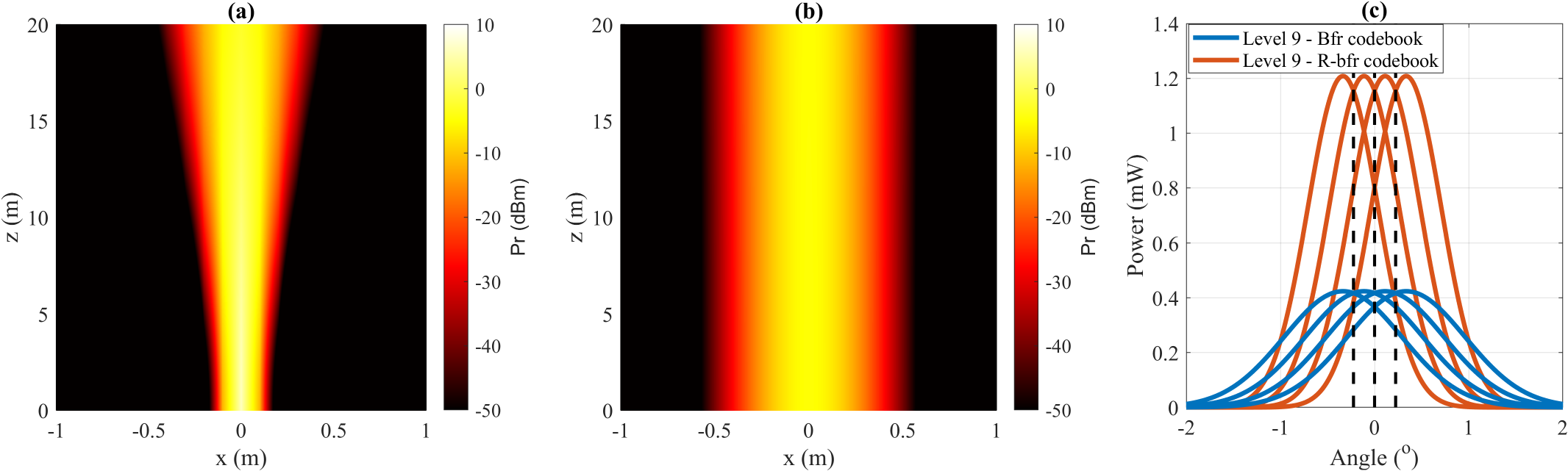}
\caption{Impact of footprint size on Phase 1 beam-forming. (a) Small ($w=7$ cm), and (b) large footprint ($w=25$ cm) on the TX. (c) Received power of four adjacent beams at 10 m from the TX with the two beam-forming codebooks. The beams are generated with the $9^{th}$ level of both codebooks, with $w=25.58$ cm for the Bfr codebook, and $w=5.41$ cm (footprint for level 7 of the Bfr codebook) for the R-bfr codebook. The dashed lines mark the directions where two adjacent beams offer the same power at the RX.}
\label{fig:fig5}
\end{figure*}
The corners in the CDFs indicate a change in behavior that happens at different localization error depending on the last beam-focusing codebook level. In Fig. \ref{fig:fig8}, the CDF of the localization error is investigated as a function of the last codebook level of the Bfc codebook in scenario 2 when in Phase 1 the Bfr codebook is used. It is evident that with higher codebook levels the corner is moved to the lower localization errors as the focal areas are smaller. Furthermore, with levels 7 and 8 the corners are at a lower probability, and also a second corner appears. The reason is that due to the small size of the focal areas, the resolution of Phase 1 becomes insufficient and a higher codebook level is required. 
Moreover, for the same reason, codebook level 7 performs better than level 8 at localization error $4$ cm. The same conclusions can be made for scenario 1.

\section{Practical considerations \& Low complexity approach} 
The success of the algorithm for achieving localization accuracy with the desired resolution depends on successfully locating the RX direction in Phase 1 and, subsequently, finding the focal area closest to the RX in Phase 2, along that direction. The direction estimated in Phase 1, however, involves an angular error of the order of the beamwidth, as shown in Fig. \ref{fig:fig5}(a). In this example the footprint is $w=w_x=w_y=7$ cm and the 3dB beamwidth is $\theta_{3dB}=\sqrt{2 \ln2}\ \lambda/(\pi w)=0.61^o$. With narrower beams, the RX direction can be estimated with better accuracy, overall reducing the localization error in Phase 2. An intuitive way to form narrower beams and increase the resolution in Phase 1 would be to increase the footprint, as shown in Fig. \ref{fig:fig5}(b). In this example, $w=25$ cm and the 3dB beamwidth is $0.17^o$. Note, however, that the beamwidth refers to the beam extent in its far-field, while the beam in Fig.6(b) is clearly still evolving in its near-field (the Rayleigh length here is $z_R=98$ m). It turns out that, because larger footprints increase the Rayleigh length and push the near-to-far field transition to farther distances, increasing the footprint beyond some point becomes impractical; for the area of interest the beams evolve entirely in their near-field and, instead of becoming narrower with larger footprint, they become wider. Moreover, because in the near-field regime, beam diffraction is weak, the beams propagate with practically constant peak power, as shown in Fig. \ref{fig:fig5}(b). As a result, the received power along the propagation distance, and among adjacent beams of the codebook, remains almost unchanged, as shown in Fig. \ref{fig:fig5}(c) where the beams of the Bfr codebook become almost indistinguishable at the RX. Therefore, it becomes difficult to differentiate between two adjacent beams of the Bfr codebook, particularly at the higher codebook levels.\\
\indent One simple solution to this problem in Phase 1 is to use the footprint of a lower codebook level with the directions of the higher levels of the codebook. This codebook will be called robust beam-forming (R-bfr) codebook, and as shown in Fig. \ref{fig:fig5}(c), it offers higher power with narrower beams than the Bfr codebook, making it more robust to AWGN. In the simulations presented next, the R-bfr codebook generates beams at the same directions as the Bfr codebook but it retains the same footprint from the $7^{th}$ level and up. This results in a footprint radius of $6.5$ cm, and a Rayleigh length of $\sim6.6$ m.
\begin{figure}
\centering
\includegraphics[width=\linewidth]{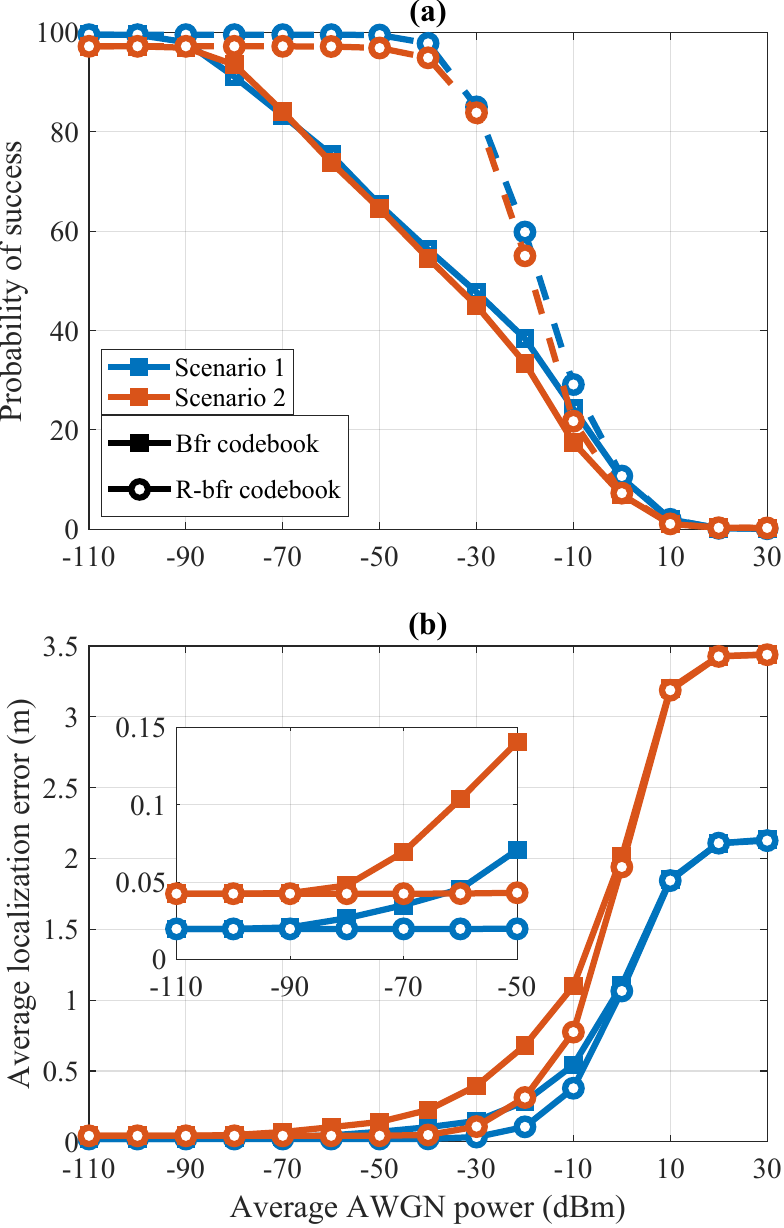}
\caption{(a) Probability of success for the two scenarios and codebooks, and (b) average localization error vs average AWGN power. For convenience, the inset shows the probability of success for average AWGN power up to $-50$ dBm.}
\label{fig:fig6}
\end{figure}
As an example, in Fig. \ref{fig:fig6}(a), the probability of success is presented for both scenarios as a function of $P_n$, the average AWGN power at the RX. 
The algorithm is considered successful when the error of the estimated location is at most equal to the resolution presented in Table \ref{tab:tab1}. The probability of success has the same behavior in both scenarios regardless of the codebook.
In both scenarios, the R-bfr codebook is significantly more robust to AWGN than the Bfr codebook. For instance, with the Bfr codebook, the probability of success starts decreasing from $P_n=-80$ dBm, while with the R-bfr codebook from $P_n= -40$ dBm. The probability of success reaches zero at $20$ dBm regardless of the scenario and codebook. Moreover, regardless of the scenario, the difference between the two codebooks increases from $P_n=-80$ dBm to $P_n=-40$ dBm where the difference is highest at $40\%$, and then it starts decreasing.
In Fig. \ref{fig:fig6}(b), the average localization error is presented, as a function of the average AWGN power. 
The minimum and maximum values for the average localization error is the same with both codebooks regardless of the scenario. In scenario 1, the minimum value is $2$ cm, and the maximum is $2.1$ m, while in scenario 2, they are $4.2$ cm, and $3.4$ m respectively.
In both scenarios, with the Bfr codebook, the average localization error is the same from $-110$ dBm to $-90$ dBm. With the R-bfr codebook, the algorithm can retain low average error for higher values of AWGN, as in both scenarios, the average error stays the same from $-110$ to $-50$ dBm (see inset in Fig. \ref{fig:fig6}(b)). The maximum value for the average error is again reached at $20$ dBm.

\begin{figure}
\centering
\includegraphics[width=\linewidth]{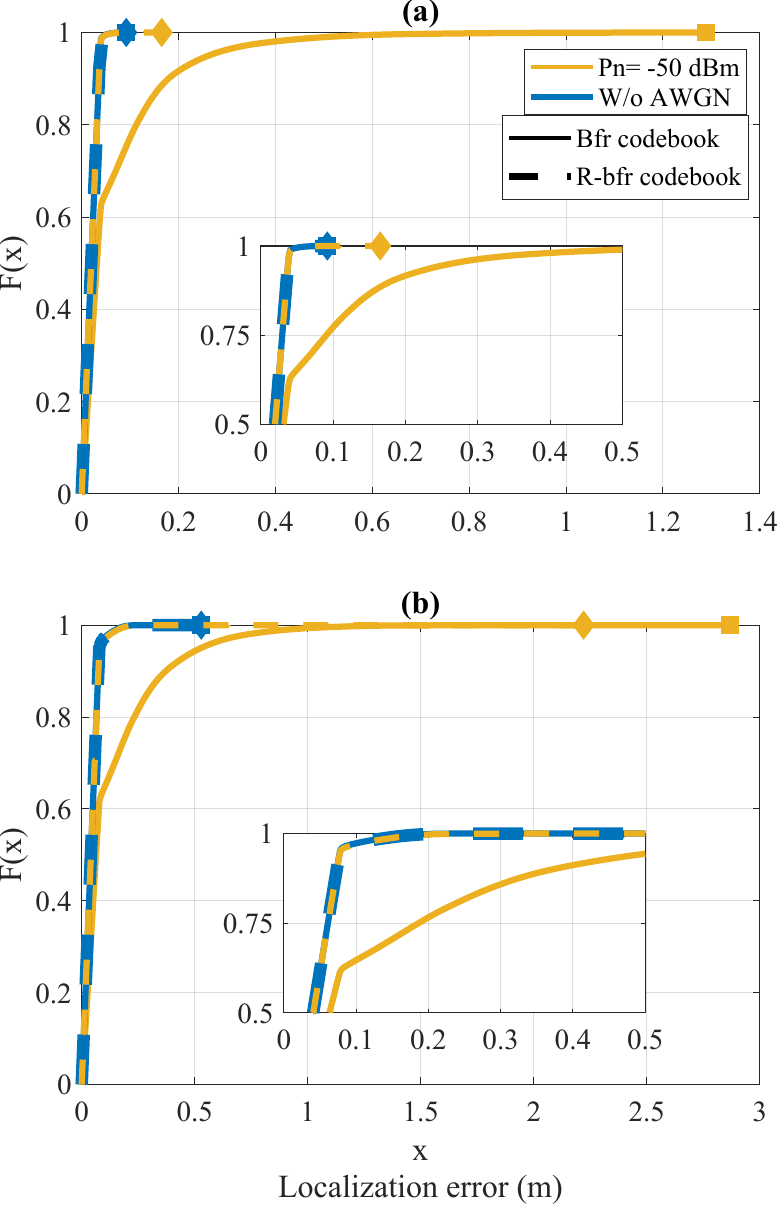}
\caption{Empirical CDF of the localization error in scenarios 1, and 2, in (a), and (b), respectively. The comparison is between the Bfr and R-bfr codebooks, with and without AWGN. The squares mark the maximum measured error of the algorithm with the Bfr codebook, and the rhombi the maximum measured error with the R-bfr codebook. For convenience, the insets show a small part of the CDFs.}
\label{fig:fig7}
\end{figure}

In Fig. \ref{fig:fig7}, the empirical CDF of the localization error is presented for the two scenarios, with and without AWGN at the RX, $P_n$, and the two codebooks mentioned previously. The performance in scenarios 1, and 2 are shown in (a), and (b), respectively. In the case with AWGN, $P_n=-50$ dBm, where the difference in probability of success between the two codebooks is $\sim 32\%$ in both scenarios as shown in Fig. \ref{fig:fig6}(a). 
As previously explained, with the Bfr codebook, the algorithm is prone to errors due to the AWGN. 
For example, in scenario 1 in (a), the overall maximum error with the Bfr codebook in the case without AWGN is $9.2$ cm, while in the case $P_n= -50$ dBm it is $1.29$ m. Furthermore, the 99.9\% maximum error without AWGN, and with $P_n= -50$ dBm is $6.6$ cm and $91$ cm, respectively, while the average localization error is $2$ cm, and $7$ cm.
With the R-bfr codebook, the overall maximum error is $9.2$ cm in the cases without AWGN, and with $P_n=-50$ dBm it is $16.5$ cm. Furthermore, the 99.9\% maximum error is $6.6$ cm, and the average localization error is $2$ cm in all cases. It can be observed that with the Bfr codebook, both the overall and the 99.9\% maximum errors increase in the presence of AWGN. With the R-bfr codebook, however, they stay mostly the same with the case without AWGN as it is more robust. 
In scenario 2 in (b), the effect of AWGN is even more pronounced due to the larger size and the lower power density of the focal areas per codebook level which is the result of the increased maximum distance of the area of interest. With the Bfr codebook, in the case without AWGN, the overall maximum error is $53$ cm, the 99.9\% maximum error is $20$ cm, and the average localization error is $4.3$ cm. 
With $P_n= -50$ dBm, the overall maximum error increases to $2.8$ m, the 99.9\% maximum error to $1.3$ m, and the average localization error to $14$ cm. 
On the other hand, with the R-bfr codebook the overall maximum error is $53$ cm, the $99.9\%$ maximum error is $21$ cm, and the average localization error is $4.2$ cm for the cases without AWGN. With $P_n=-50$ dBm, the overall maximum error is $2.2$ m, the $99.9\%$ maximum error is $24$ cm, and the average localization error is $4.3$ cm. 

\section{Mobile user tracking}
\begin{figure}
\centering
\includegraphics[width=\linewidth]{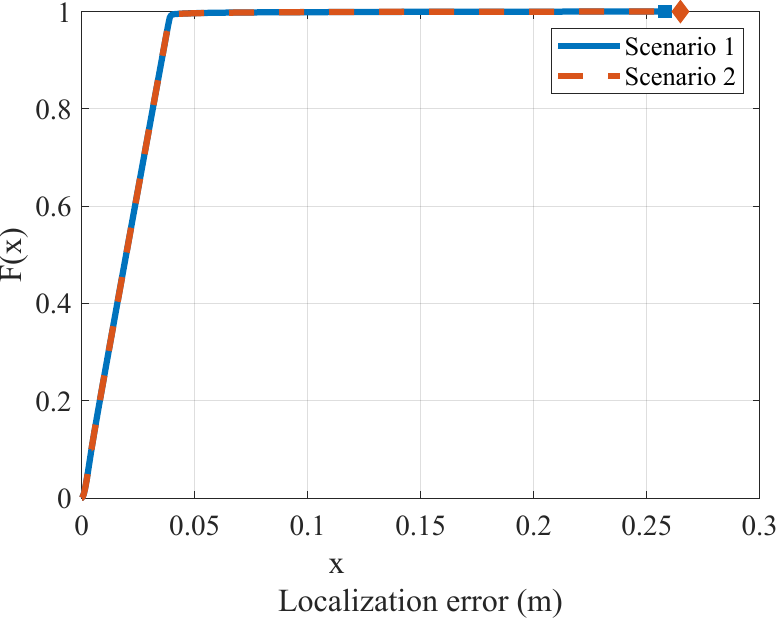}
\caption{Empirical CDF of the localization error from location tracking of mobile users in scenarios 1, and 2. The markers show the maximum error in each scenario.}
\label{fig:fig11}
\end{figure}

Tracking the location of a mobile user is different than locating a static user as exhaustive search or even the codebook based search shown in Section \ref{s:sec4} can be time consuming depending on the speed of the user. Therefore, the mobility, and especially the speed, of the user must be taken into account. In this section, the localization algorithm in Section \ref{s:sec4} is adapted in order to track the location of mobile users. In particular, the new algorithm predicts the future location of the user by taking into account the distance covered in three timeslots assuming linear motion and fixed speed. Then in Phase 1, starting with the $(L_t-3)$-th codebook level, the algorithm scans four directions around the predicted location. In Phase 2, the algorithm scans four focal areas with the highest codebook level. As a result, the overhead is significantly reduced compared to the localization of static users. Furthermore, since the areas of interest are known to the TX, the predicted locations that fall outside of them, are adjusted angle-wise to be inside of them. For example, if the angle of a predicted location is $27^o$, while the maximum angle in the area of interest is $25^o$, the algorithm uses the beams that are close to $25^o$. 
The performance is evaluated through Monte-Carlo simulations with $10^6$ iterations, where the user follows random linear trajectories inside the areas of interest of scenarios 1, and 2, shown in Fig. \ref{fig:figsc}, with the minimum distance from the TX being $10$ cm. 
Moreover, the minimum distance that each motion covers is $3$ m. In scenario 1, the maximum beam-forming codebook level is $L_t= 10$, and the the maximum beam-focusing codebook level is $L_r= 6$, while in scenario 2, they are $L_t= 11$, and $L_r= 7$, respectively.
It is observed that the performance of the tracking algorithm is almost the same in the two scenarios. The maximum error in scenario 2 is $26.5$ cm, which is slightly higher than in scenario 1, where it is $25.6$ cm. In contrast, the maximum error of scenario 1 with static users is $9.2$ cm, while in scenario 2 it is $53$ cm. Furthermore, in Fig. \ref{fig:figsc}, the $99.9\%$ maximum error is $3.9$ cm, and the average error is $2$ cm in both scenarios, which are almost the same as the ideal case of scenario 1 shown in Section \ref{s:sec4}. The reason for the low $99.9\%$ and average errors is that the algorithm uses only the last codebook level in Phase 2 that generates the smallest focal areas whose size is unaffected when $\theta_r\neq 0$.

\section{Conclusions}
In this work, a localization algorithm was presented that can be implemented with LAAs and LISs. The algorithm offers low average localization error, which, depending on the AWGN power levels, and the maximum distance in the area of interest from the TX, is $2$ cm, or $4.2$ cm. Furthermore, an hierarchical ranging algorithm that is the equivalent of hierarchical beam-training was proposed, and the codebook generation method was presented. The localization algorithm consists of two phases, hierarchical beam-training and the proposed hierarchical ranging. Both beam-forming and beam-focusing can be used in communications. Therefore the ranging and localization algorithms can be easily integrated into a communications system that makes use of both techniques, making them suitable for ISAC and a communicate-to-sense approach. Furthermore, a simple approach for the beam-forming codebook was presented to make it more robust to AWGN. 
The performance of the localization algorithm was presented through the probability of success, average localization error, and empirical CDFs, with and without the presence of AWGN. Both show that the performance of  the algorithm depends on the resolution of Phase 1 being comparable to the resolution of Phase 2. Moreover, the R-bfr codebook makes the algorithm considerably more robust to AWGN. Finally, a location tracking algorithm was also presented  and evaluated with the empirical CDF, with higher accuracy in scenarios with long distances from the TX and lower overhead than the localization for static users.

\section*{Acknowledgment}
\noindent This work was supported by the European Commission’s Horizon Europe Programme under the Smart Networks and Services Joint Undertaking TERA6G project (Grant Agreement 101096949) and INSTINCT project (Grant Agreement 101139161).

\bibliographystyle{IEEEtran}
\bibliography{IEEEabrv,References}

\begin{thebibliography}{10}
\providecommand{\url}[1]{#1}
\csname url@samestyle\endcsname
\providecommand{\newblock}{\relax}
\providecommand{\bibinfo}[2]{#2}
\providecommand{\BIBentrySTDinterwordspacing}{\spaceskip=0pt\relax}
\providecommand{\BIBentryALTinterwordstretchfactor}{4}
\providecommand{\BIBentryALTinterwordspacing}{\spaceskip=\fontdimen2\font plus
\BIBentryALTinterwordstretchfactor\fontdimen3\font minus
  \fontdimen4\font\relax}
\providecommand{\BIBforeignlanguage}[2]{{%
\expandafter\ifx\csname l@#1\endcsname\relax
\typeout{** WARNING: IEEEtran.bst: No hyphenation pattern has been}%
\typeout{** loaded for the language `#1'. Using the pattern for}%
\typeout{** the default language instead.}%
\else
\language=\csname l@#1\endcsname
\fi
#2}}
\providecommand{\BIBdecl}{\relax}
\BIBdecl

\bibitem{Chaccour2022}
C.~Chaccour, M.~N. Soorki, W.~Saad, M.~Bennis, P.~Popovski, and M.~Debbah,
  ``Seven defining features of terahertz (thz) wireless systems: A fellowship
  of communication and sensing,'' \emph{IEEE Communications Surveys \&
  Tutorials}, vol.~24, no.~2, pp. 967--993, 2022.

\bibitem{Jiang2024}
W.~Jiang, Q.~Zhou, J.~He, M.~A. Habibi, S.~Melnyk, M.~El-Absi, B.~Han, M.~D.
  Renzo, H.~D. Schotten, F.-L. Luo, T.~S. El-Bawab, M.~Juntti, M.~Debbah, and
  V.~C.~M. Leung, ``Terahertz communications and sensing for 6g and beyond: A
  comprehensive review,'' \emph{IEEE Communications Surveys \& Tutorials}, pp.
  1--1, 2024.

\bibitem{Balzer2022}
J.~C. Balzer, C.~J. Saraceno, M.~Koch, P.~Kaurav, U.~R. Pfeiffer,
  W.~Withayachumnankul, T.~Kürner, A.~Stöhr, M.~El-Absi, A.~A.-H. Abbas,
  T.~Kaiser, and A.~Czylwik, ``Thz systems exploiting photonics and
  communications technologies,'' \emph{IEEE Journal of Microwaves}, vol.~3,
  no.~1, pp. 268--288, 2023.

\bibitem{Galeote2024}
J.~E. Galeote-Cazorla, A.~Ramírez-Arroyo, S.~Moreno-Rodríguez, J.-M.
  Molina-García-Pardo, M.-T. Martinez-Inglés, P.~Padilla, and J.~F.
  Valenzuela-Valdés, ``A study on w-band frequency attenuation in the presence
  of human blockage,'' in \emph{2024 18th European Conference on Antennas and
  Propagation (EuCAP)}, 2024, pp. 1--5.

\bibitem{Petrov2019}
V.~Petrov, D.~Moltchanov, S.~Andreev, and R.~W. Heath, ``Analysis of
  intelligent vehicular relaying in urban 5g+ millimeter-wave cellular
  deployments,'' in \emph{2019 IEEE Global Communications Conference
  (GLOBECOM)}, 2019, pp. 1--6.

\bibitem{Stratidakis2020}
G.~Stratidakis, E.~N. Papasotiriou, H.~Konstantinis, A.-A.~A. Boulogeorgos, and
  A.~Alexiou, ``Relay-based blockage and antenna misalignment mitigation in
  {THz} wireless communications,'' in \emph{2020 2nd 6G Wireless Summit (6G
  {SUMMIT})}.\hskip 1em plus 0.5em minus 0.4em\relax {IEEE}, Mar 2020.

\bibitem{Teoman2019}
E.~Teoman and T.~Ovatman, ``Trilateration in indoor positioning with an
  uncertain reference point,'' in \emph{2019 IEEE 16th International Conference
  on Networking, Sensing and Control (ICNSC)}, 2019, pp. 397--402.

\bibitem{Yassin2017}
A.~Yassin, Y.~Nasser, M.~Awad, A.~Al-Dubai, R.~Liu, C.~Yuen, R.~Raulefs, and
  E.~Aboutanios, ``Recent advances in indoor localization: A survey on
  theoretical approaches and applications,'' \emph{IEEE Communications Surveys
  \& Tutorials}, vol.~19, no.~2, pp. 1327--1346, 2017.

\bibitem{Chen2022a}
H.~Chen, H.~Sarieddeen, T.~Ballal, H.~Wymeersch, M.-S. Alouini, and T.~Y.
  Al-Naffouri, ``A tutorial on terahertz-band localization for 6g communication
  systems,'' \emph{IEEE Communications Surveys \& Tutorials}, vol.~24, no.~3,
  pp. 1780--1815, 2022.

\bibitem{Nagy2020}
A.~Nagy, T.~Bigler, A.~Treytl, R.~Stenzl, S.~Wilker, T.~Sauter, and T.~Wien,
  ``{RSS}-based localization for directional antennas,'' in \emph{2020 25th
  IEEE International Conference on Emerging Technologies and Factory Automation
  (ETFA)}, vol.~1, 2020, pp. 774--781.

\bibitem{Malhotra2005}
N.~Malhotra, M.~Krasniewski, C.~Yang, S.~Bagchi, and W.~Chappell, ``Location
  estimation in {Ad Hoc} networks with directional antennas,'' in \emph{25th
  IEEE International Conference on Distributed Computing Systems (ICDCS'05)},
  June 2005, pp. 633--642.

\bibitem{Wei2023}
Z.~Wei, H.~Qu, Y.~Wang, X.~Yuan, H.~Wu, Y.~Du, K.~Han, N.~Zhang, and Z.~Feng,
  ``Integrated sensing and communication signals toward 5g-a and 6g: A
  survey,'' \emph{IEEE Internet of Things Journal}, vol.~10, no.~13, pp.
  11\,068--11\,092, 2023.

\bibitem{Pin2021}
D.~K. Pin~Tan, J.~He, Y.~Li, A.~Bayesteh, Y.~Chen, P.~Zhu, and W.~Tong,
  ``Integrated sensing and communication in 6g: Motivations, use cases,
  requirements, challenges and future directions,'' in \emph{2021 1st IEEE
  International Online Symposium on Joint Communications \& Sensing (JC\&S)},
  2021, pp. 1--6.

\bibitem{INSTINCT}
``Instinct,'' \url{https://www.barkhauseninstitut.org/en/instinct-}\\
  \url{joint-sensing-and-communications-for-future-connectivity}.

\bibitem{Huang2023}
Y.~Huang, J.~Yang, W.~Tang, C.-K. Wen, S.~Xia, and S.~Jin, ``Joint localization
  and environment sensing by harnessing nlos components in {RIS}-aided mmwave
  communication systems,'' \emph{IEEE Transactions on Wireless Communications},
  vol.~22, no.~12, pp. 8797--8813, Dec 2023.

\bibitem{Lu2024}
Y.~Lu, Z.~Zhang, and L.~Dai, ``Hierarchical beam training for extremely
  large-scale mimo: From far-field to near-field,'' \emph{IEEE Transactions on
  Communications}, vol.~72, no.~4, pp. 2247--2259, April 2024.

\bibitem{Que2023}
H.~Que, J.~Yang, C.-K. Wen, S.~Xia, X.~Li, and S.~Jin, ``Joint beam management
  and slam for mmwave communication systems,'' \emph{IEEE Transactions on
  Communications}, vol.~71, no.~10, pp. 6162--6179, 2023.

\bibitem{El-Absi2018}
M.~El-Absi, A.~Alhaj~Abbas, A.~Abuelhaija, F.~Zheng, K.~Solbach, and T.~Kaiser,
  ``High-accuracy indoor localization based on chipless rfid systems at thz
  band,'' \emph{IEEE Access}, vol.~6, pp. 54\,355--54\,368, 2018.

\bibitem{Fan2021}
S.~Fan, Y.~Wu, C.~Han, and X.~Wang, ``Siabr: A structured intra-attention
  bidirectional recurrent deep learning method for ultra-accurate terahertz
  indoor localization,'' \emph{IEEE Journal on Selected Areas in
  Communications}, vol.~39, no.~7, pp. 2226--2240, July 2021.

\bibitem{Kanhere2020}
O.~Kanhere and T.~S. Rappaport, ``Millimeter wave position location using
  multipath differentiation for 3gpp using field measurements,'' in
  \emph{GLOBECOM 2020 - 2020 IEEE Global Communications Conference}, Dec 2020,
  pp. 1--7.

\bibitem{Wymeersch2021}
H.~Wymeersch, D.~Shrestha, C.~M. de~Lima, V.~Yajnanarayana, B.~Richerzhagen,
  M.~F. Keskin, K.~Schindhelm, A.~Ramirez, A.~Wolfgang, M.~F. de~Guzman,
  K.~Haneda, T.~Svensson, R.~Baldemair, and S.~Parkvall, ``Integration of
  communication and sensing in 6g: a joint industrial and academic
  perspective,'' in \emph{2021 IEEE 32nd Annual International Symposium on
  Personal, Indoor and Mobile Radio Communications (PIMRC)}, Sep. 2021, pp.
  1--7.

\bibitem{Rao2024}
C.~Rao, Z.~Ding, O.~A. Dobre, and X.~Dai, ``A general analytical framework for
  the resolution of near-field beamforming,'' \emph{IEEE Communications
  Letters}, vol.~28, no.~5, pp. 1171--1175, May 2024.

\bibitem{Wu2024}
Z.~Wu, M.~Cui, and L.~Dai, ``Enabling more users to benefit from near-field
  communications: From linear to circular array,'' \emph{IEEE Transactions on
  Wireless Communications}, vol.~23, no.~4, pp. 3735--3748, April 2024.

\bibitem{Droulias2024}
S.~Droulias, G.~Stratidakis, and A.~Alexiou, ``Near-field engineering in
  {RIS}-aided links: Beamfocusing analytical performance assessment,''
  \emph{IEEE Access}, vol.~12, pp. 29\,536--29\,546, 2024.

\bibitem{Droulias2022}
S.~Droulias and A.~Alexiou, ``Reconfigurable intelligent surface: an angular
  spectrum representation approach,'' in \emph{Asilomar Conference on Signals,
  Systems, and Computers}, 2022.

\bibitem{Wu2020}
Y.~Wu, B.~Jiang, M.~Zhang, J.~Hirokawa, and Q.~H. Liu, ``A four-corner-fed
  slotted waveguide sparse array for near-field focusing,'' \emph{IEEE Access},
  vol.~8, pp. 203\,048--203\,057, 2020.

\bibitem{Stratidakis2024}
G.~Stratidakis, S.~Droulias, and A.~Alexiou, ``Beam-tracking challenges in
  {THz} communications,'' \emph{Wireless World Research and Trends Magazine},
  pp. 13--20, Mar. 2024.

\bibitem{Sayeed2015}
A.~Sayeed, ``Beamspace mimo architectures for massive 2{D} antenna arrays,''
  plenary presentation at the International Workshop on Emerging MIMO
  Technologies with 2D Antenna Arrays,ICNC 2015, Feb. 2015.

\end{thebibliography}

\end{document}